\documentclass[twocolumn]{imsart}

\usepackage{comment}
\RequirePackage{amsthm,amsmath,amsfonts,amssymb}
\RequirePackage[numbers]{natbib}
\RequirePackage{graphicx}

\startlocaldefs

\endlocaldefs

\begin{document}

\newtheorem{definition}{Definition}[section]
\newcommand{\etal}{{\it et al.\ }}

\begin{frontmatter}
\hrulefill

\title{Defining Replicability of Prediction Rules} 
\runtitle{Replicability of Predictions}

\begin{aug}
\author[A]{\fnms{Giovanni}~\snm{Parmigiani}\ead[label=e1]{gp@ds.dfci.harvard.edu}}
\address[A]{Giovanni Parmigiani is Professor, Department of Data Science, Dana Farber Cancer Institute \& Department of Biostatistics, Harvard T.H. Chan School of Public Health\printead[presep={\ }]{e1}.}
\end{aug}

\hrulefill
\begin{center}
    Draft, \today
\end{center}

\begin{abstract}
In this article I propose an approach for defining replicability for prediction rules. 
Motivated by a recent NAS report, I start from the perspective that replicability is obtaining consistent results across studies suitable to address the same prediction question, each of which has obtained its own data. I then discuss concept and issues in defining key elements of this statement. I focus specifically on the meaning of "consistent results" in typical utilization contexts, and propose a multi-agent framework for defining replicability, in which agents are neither partners nor adversaries. I recover some of the prevalent practical approaches as special cases. I hope to provide guidance for a more systematic assessment of replicability in machine learning.
\end{abstract}

\begin{keyword}
\kwd{Replicability}
\kwd{Prediction}
\kwd{Decision Theory}
\vspace{12pt}

\end{keyword}

\end{frontmatter}


\section{Introduction}

\subsection{Preface}

Prediction and machine learning technologies are playing increasingly important roles in science, as many fields leverage rapidly evolving data-generating technologies. Yet replicability, to many an essential element of science, remains inadequately studied in prediction, in part because its definition in relation to prediction remains somewhat elusive. In this article, I discuss concepts and issues in replicability of prediction from a perspective rooted in my experience as a practitioner of prediction approaches in biomedical research, and propose a framework for defining replicability.

\subsection{Examples} \label{sec:ex}

I begin with examples to give a concrete sense of application contexts and constituents. All are drawn from medicine, where prediction rules are regularly used to support decision making, an replicability is a practical concern. Based on my much more limited experience I hope the framework I propose will also guide investigations in many other areas. While illustrative examples may make the concepts more concrete, my intent is not to provide a descriptive account, but rather to encourage discussion about prescriptive theories of replicability quantification. 

In medicine, recent years have seen a trend toward regulating models and software using criteria similar to those used for medical devices. For example, the European Commission’s Medical Devices Coordination Group has published guidance on the classification of software for regulatory purposes. A critical element of their definition is whether the software is intended to support decision making affecting patients or the public \cite{Beckers2021}. Regulatory requirements include validation in independent data as well as surveillance of performance in newly gathered data from practical settings after approval. These general trends, as well as many specific applications, are motivating the definitions in this paper.

\subsubsection{Predicting Sepsis.} \label{ex:um}
Models alerting clinicians to the presence of bacterial sepsis are widely used in emergency medicine. In April 2020, one such model, provided by a commercial entity (Epic Systems), was deactivated by one of the hospitals using it (the University of Michigan Hospital) because of the frequency of false positive alerting associated with COVID-19 \cite{Finlayson2021nejm}. To investigate this issue, \cite{Wong2021hi}, quantified the performance of sepsis detection models in 24 hospitals before and during the COVID-19 pandemic. Among individuals with the coronavirus, the relationship between fevers and sepsis differs from what is observed in the majority of the individuals in the dataset originally used for training the model. This leads to a deterioration of the performance of the model, as captured for example by an increase in the frequency of alerts, as the prevalence of individuals with COVID-19 increases. In this case the issue emerged as a result of a data collection specific to the user's context. More broadly, Kelly \etal \cite{Kelly2019mbcm} discuss challenges for clinical implementation of machine learning algorithms across a diverse set of populations and systems. 

\subsubsection{Predicting Survival of Ovarian Cancer Patients.} \label{ex:oc}
High throughput measurement of gene transcription offers an opportunity to more accurately predict the survival times of patients diagnosed with cancer. In Waldron \etal \cite{Waldron2014}, focusing on ovarian cancer, we carried out a comprehensive review of published prediction / scoring rules of this kind, identifying 14 from published articles that could be re-coded with a high degree of reproducibility. Modelers divulge these algorithms with clinical users in mind, hoping they will serve to inform decision making at the bedside. In parallel, we comprehensively surveyed and collected all available data sets that could be used to assess model performance and systematically evaluated every rule on every dataset. Our work followed, and was in part motivated by, a case involving "premature use of omics-based tests" \cite{InstituteofMedicine:vr}. Our results, among others such as Chang and Geman \cite{Chang:2015ua}, systematically document a consistent gap between the performance of prediction rules within the training studies (as measured, say, by cross-validation) and the performance of the same rule in relevant independent datasets. I will revisit this example later as "the ovarian cancer example".

\subsubsection{Evaluating Retinal Images.} \label{ex:va} 
Diabetic retinopathy is diagnosed with the support of imaging techniques. Automated interpretation of images is important for primary care settings. Investigators at the United States' Veteran Administration (VA) Health System \cite{Lee2021dc} carried out a large prospective multi-center validation study to perform a head-to-head comparison of seven algorithms, including one FDA-approved algorithm, evaluating retinal images. I will revisit this example later as "the VA example". 

\subsubsection{Screening for Tuberculosis.} \label{ex:tb}
Chest radiography is used to screen people for pulmonary tuberculosis (TB). Deep learning (DL) neural networks are now available to interpret the images. \citep{Qin2019sr} acquired images from two existing studies in Nepal and Cameroon, and compared three commercially available deep learning neural networks algorithms in both countries. I will revisit this example later as "the TB example".  

\subsubsection{External Assessment.} In each of this examples, replicability is evaluated via multiple data sets, with either no overlap of individual units with the data used to train the prediction rule, or a clear indication of whether this overlap exists and how it affects the results.

\section{Glossary}

In this section I try to clarify the use of the terms "prediction", "prediction rule", "replicability" and "study", also pointing briefly to challenges and issues with these definitions.

\subsection{Prediction} \label{sec:pred}

A prediction, for the purpose of this discussion, is a statement $p \in \cal P$ about a
future or unknown observable $y \in \cal Y$ (the label). A prediction rule generates predictions on the basis of observations $x \in \cal X$ (the predictors), and is thus a mapping $\phi : \cal X \rightarrow \cal P$.
In scoring systems $\cal P \subseteq \mathbb{R}$; in statistical prediction $\cal P$ is either a probability space on $\cal Y$ or a probability space on probability distributions on $\cal Y$. I will also consider the simple binary case where the algorithm directly assigns each point to one of two possible classes, in which case $\cal P = \cal Y$.

The terms {\it observable} and {\it unknown}, which I used in my definition, are far from being self-explanatory.  By {\it observed} ($y$ or $x$) I mean that there is agreement, within a relevant group of individuals do be discussed further, about the precise value of the labels or predictors. This is not to say that I exclude disagreement altogether. Say radiologists A and B classify the label of the same medical image differently. Radiologist A judges it to reveal a "malignant" condition while B judges it to be "benign". This could be formalized by defining separate dimensions $y_A$ and $y_B$ within $y$. My discussion, however, is within confines where at some point, the disagreement within the group ceases, for example because it is at least agreed that the radiologists' answers are indeed $y_A$ and $y_B$.  By {\it observable} I mean that, should the observation be performed, there will typically be agreement on the value of the result. By "unknown" $y$, I mean simply that knowledge of the value of $y$ is not part of the making of~$\phi$.

More broadly, observations are not in general separable from the theories that provided the framework to generate them, and from the contextual values of the prediction tasks. Think of predicting an individual's mental health outcomes, or their subversive political behavior, as examples. Thus my definitions, and by extension any ensuing consideration about replicability, are contextual to the goals, value systems, and theories that underlie the agreement among the individuals in the reference group. The nature and size of the reference group may vary widely in different contexts.

Prediction is regarded by several pioneers of statistical thought as the fundamental problem of statistics \cite{Geisser2014wol}. Examples include de~Finetti, Pearson and arguably Bayes \cite{Stigler1982jrssa} and Laplace. Motivations for favoring predictive approaches include both empiricist and pragmatist considerations. An important motivation when discussing foundations is avoiding the additional degree of abstraction necessary to define concepts such as parameter, hypothesis, representation, latent class and so forth.

\subsection{Replicability} \label{sec:repl}

Among the fundamental premises of the scientific enterprise is a degree of concordance among experimental observations made in sufficiently similar circumstances. From this follows the desideratum that scientific predictions also agree well with experimental observations made across sufficiently similar circumstances. Defining this rigorously is not straightforward.

In 2019 the USA's National Academies established a Committee on Reproducibility and Replicability in Science. Their report \cite{CommitteeonReproducibilityandReplicabilityinScience:2019hx} is essential reading for those interested in this topic. Though their focus is primarily on scientific hypotheses, their definition is a good starting point for this discussion. Conclusion 3-1 on page 36 states: "Replicability is obtaining consistent results across studies aimed at answering the same scientific question, each of which has obtained its own data". This definition is consistent with that given by the American Statistical Association \cite{Association:98z9A4lW} and with other work in statistics \cite{Heller:2014cn}. 

A distinct concept is that of repeatability: a repeatable prediction approach produces predictions without variation across independent tests carried out by repeating the entire process, including data collection, on the same individual or sampling unit \cite{Lemay2022njpdm}. This is important but is not examined here.

Replicability is also used in contrast to reproducibility, defined as "obtaining consistent results using the same input data, computational steps, methods, and code, and conditions of analysis" \cite{CommitteeonReproducibilityandReplicabilityinScience:2019hx}. Usage of these terms is often inconsistent and is plainly reversed in computer science ---a sarcastic twist in the parallel evolution of the same concepts in siloed fields. A thread of literature debates and documents related terminologies and their usage
\cite{Kenett:2015kb,Goodman2016stm,Barba2018a}.

For predictions, I propose to modify the NAS definition to say: 
\begin{definition}[Replicability of Prediction Rules] \label{def:rep} \hspace{30pt}\\
    Replicability is obtaining consistent results across studies {\bf suitable} to address the same scientific {\bf prediction} question, each of which has obtained its own data.
\end{definition}
Key edits are in bold. I narrowed the scientific question to prediction, but I broadened the definition to the consideration of suitable studies or datasets irrespective of the original aim of the data collection or design.

Determining whether any two studies are suitable for answering the same scientific prediction question is a matter of judgment. As was the case for observations in Section~\ref{sec:pred}, my view is that it is useful to frame this determination in terms of inter-subjective agreement. Given the potential ambiguities of defining objective experimental results, and objective suitability of studies to a specific prediction question, I do not think it is useful to attempt an objective definition of prediction replicability. A more realistic goal is to aim for a broad consensus on observables and data, so that data can be used to descriptively quantify replicability in a way that will be found to be convincing by many.

The NAS report proposes to think of replication as ``the act of repeating an entire study, independently of the original investigator without the use of original data.'' A strength of such replication activity is that similarities and differences between the original study and its replica are themselves part of the experimental design. In my opinion this type of activity can in principle lead to the most compelling evidence about replicability. I will call it replication {\it by design}.

Definition~\ref{def:rep}, however, allows for a broader empirical scope, including data generation activities that may or may not originate to answer the same prediction question, or any prediction question. I will call it {\it observational} replication. While replication by design is defined in reference to a specific study and the activity of replicating it, observational replication is defined in reference to a specific prediction task and a collection of relevant datasets. Interest in forming collections of datasets for the purpose of understanding prediction rule replicability and study heterogeneity "in the wild" is growing ---see for example the WILDS data collections \cite{Koh2020pmlr}.

External validation studies of prediction rules gather evidence about the applicability of a prediction rule beyond the conditions wherein it was formulated or trained, using available independent data  \cite{Steyerberg2014ehj}. This can be implemented by design or observationally or both depending on the circumstances. There is an element of replicability in these analyses, insofar as they compare properties of prediction rules across datasets. Many study designs and analytic techniques are relevant for both tasks. An important distinction is in the questions asked. Simplifying, validation asks weather a model's prediction ability is adequate for a certain set of tasks, while replicability asks whether prediction ability varies across multiple independent studies.

\subsection{Studies} \label{sec:stu}

Formally, a study $S$ is a collection of units, where a unit is a point in $( \cal X \times \cal Y )$. So a study of size $n$ is a point in $( {\cal X} \times {\cal Y} ) ^n$. It is useful to frame discussions of replicability of predictions around a collection of relevant studies $S_1, \ldots, S_K$.
The size of study $k$ is $n_k$. 

In an example of replicability by design, focus may be on decision rule $\phi$, associated with a specific publication or software tool. Investigators may prospectively perform replication studies $S_1, \ldots, S_K$, (as in Example~\ref{ex:va} where $K=2$) not necessarily from identically distributed populations. In this case we have a sharp pre-existing definition of $\phi$, $\cal X$ and $\cal Y$. 

In an example of observational replicability we may gather evidence about the applicability of $\phi$ beyond the conditions where it was trained, using existing data. $S_1, \ldots, S_K$ are chosen based on a set of inclusion criteria which could include: sufficient similarity of $\cal X$ and $\cal Y$ to those used in $\phi$; sufficient relevance of the units sampled; sufficient quality, and so forth. 
Specifics will be heavily dependent on the context so it is difficult to provide general guidance. In Section~\ref{sec:ex-stu} I will revisit the examples of Section~\ref{sec:ex} to help fix ideas. 

Generally speaking, replicability by design is implemented prospectively, while observational replicability can also occur via retrospective data collections. 
The Medical AI Evaluation Database \cite{Wu2021nm} catalogues medical artificial intelligence devices recently approved by the United States' FDA, and systematically reports on how they were evaluated before approval. Included are the three algorithms for the analysis of retinal images covered in Example~\ref{ex:va}, where prospective validation studies were carried out after the training of the algorithm was finalized. In two cases in the database the prospective study was multi-site. 
Prospective validation now accounts for a small minority of the approval processes reported by \cite{Wu2021nm}, just 4 of 130. However, there is interest in a more systematic use of validation by design via prospective studies~\cite{Ebrahimian2022ar}.

\subsection{Study-to-Study Variability} \label{sec:ssv}

A useful way to think about replicability is to identify interesting sources of variation across which it would be desirable for $\phi$ to be replicable, and define studies accordingly. For example, these can include variation in the technologies used for data collection, or in the selection criteria for including study units. 

Ideally, identification of these sources of variation may begin as part of the initial study, through substantive insight as well as formal statistical analysis. Guidance on how to assess and report potential sources of variation exists in various application niches, such as the analysis of batch effects in high throughput biology \cite{Leek2008,Leek2010nrg,Zhang2020b}.


While replicability can be evaluated across any collection of studies, the utility of this assessment if far greater if the study collection is defined and gathered in a systematic and comprehensive way, and based on criteria defined prior to the replicability analysis. Considerations are similar to those relevant in meta-analysis. In Waldron {\it et al.} \cite{Waldron2014} we illustrate a meta-analytic approach to forming comprehensive data collections relevant for a specific prediction task.

One way to conceptualize $S_1, \ldots, S_K$ is to think of it as a draw from a multi-level probability model, composed by a $q_k(x,y)$ that generates units within a study, and a $q(1, \ldots,K)$ drawing study indices from a hypothetical population of studies.
Much of the relevant variability discussed so far will translate into variation in the joint distributions $q_k$.

In machine learning, cross-study heterogeneity is described as ``dataset shift''. More specifically, ``concept shift'' refers to changes in the conditional probability of labels given predictors, while ``covariate shift,'' refers to changes in the joint distribution of the predictors \citep{Kouw2019a, Zhuang2020a} and ``label shift'' refers to changes in the marginal distribtion of labels. 
Moreno-Torres \etal~\cite{MorenoTorres2012pr} review and compare terminology and concepts.

\subsection{Examples} \label{sec:ex-stu}

\subsubsection{Ovarian Cancer Example.} \label{ex:oc2}
In \citep{Waldron2014} we discuss in detail a case study where we form a collection of studies deemed suitable for the replicability analysis of a family of prognostic rules. We carried out a comprehensive review of available data, with pre-defined inclusion criteria. Sources of variation across studies include different microarray analysis technologies, differences in laboratory utilization of these technologies, differences in patient populations, including variation in stage and tumor size, and differences in clinical annotations (e.g. surgical outcomes). Nonetheless, studies are sufficiently comparable that meta-analytic biomarker discovery and model training provide robust results \cite{Riester:2014ga,Ganzfried2013}.

When data collection technologies vary across studies, a nontrivial step, both practically and conceptually, is to map variables across studies. In this example, Ganzfried \cite{Ganzfried2013} illustrates the challenges of mapping transcriptomics data across high throughput technologies.

\subsubsection{VA Example.}\label{sec:va2}
\citep{Lee2021dc} prospectively collected data within the VA system at two separate locations, which constitute the studies in this case. Studies are homogeneous in important ways, including a shared IT infrastructure and data dictionaries, but vary in the populations served and some of the clinical workflows.

\subsubsection{TB Example.} \label{ex:tb2}
\citep{Qin2019sr} retrospectively identified two relevant existing studies with sufficiently similar chest radiography images and clinical annotations. Variation arises from differences in populations and referral patterns among others. 

\section{A multi-agent framework for Replicability} \label{sec:acr}

Definition~\ref{def:rep} refers to obtaining "consistent results". In this section I propose a framework for quantifying consistency of results. 

\subsection{Roles} \label{sec:roles}

In the applications of Section~\ref{sec:ex}, obtaining "consistent results" depends on modeling strategies of the developer, utilization patterns by the user, as well as the nature and variety of the collection of studies used for assessment. The process is complex. Multiple constituencies are involved and their goals are only partly overlapping. I propose to model it by defining three essential roles. 

\begin{itemize}
    \item[*] {\it Modeler.} This is the entity developing $\phi$ through statistical / machine learning techniques. 
    \item[*] {\it User.} This is the entity applying $\phi$ to execute policy, commercial, legal, medical, or other decisions in practice.
    \item[*] {\it Assessor.} This is the entity or group defining the relevant collection of studies $S_1, \ldots, S_K$, including definition of variables, and criteria for data quality. In Section~\ref{sec:pred}, I mentioned a reference group  needs to agree on observables. The assessor groups needs to be a subset of this group. 
\end{itemize}

My premise is that it is helpful to distinguish these three roles to arrive at definitions that address important societal uses of prediction algorithms. In some applications, roles may overlap. For example the user and assessor may be the same entity. In others, the developer may also be the user. I find it hard, however, to frame replicability as a traditional single-agent decision problem in the vein of, say Savage \cite{sava:1954}. In none of the examples of Section~\ref{sec:ex} are the interests of all entities involved fully aligned, although broadly speaking all agents may have an interest in replicability to occur.

In one version of these roles, the modeler, solely concerned about the construction of a useful $\phi$, is {\it Algorithmic} in the sense of Breiman's two cultures \cite{Breiman2001}; the user, immersed in a specific medical or commercial reality with clearly defined goals, is a {\it Rational Bayesian} in the tradition of Ramsey \cite{rams:1926}; and the assessors, in an effort towards neutrality, limit their scope to {\it Descriptive} statistics, a practice as old as the field. Alternatively, assessors could take a Fisherian perspective \cite{Fisher1925} and test for significance of departures from replicability. More on this in Section~\ref{sec:rej}.

Throughout, I assume that the modeler has not used any of studies $S_1, \ldots, S_K$ in the training of $\phi$.

\subsection{Single User} \label{sec:dtr}

Consider first the scenario where $\phi$ is used by a single rational agent for supporting a specific decision, defined as the choice of a point $a$ in a decision space $\cal A$ with the goal of maximizing the expectation of a utility function 
$$
U (a, x, y) : ( \cal A \times \cal X \times \cal Y) \rightarrow \mathbb{R}.
$$
A decision function is a mapping $\delta (\phi) : {\cal P} \rightarrow {\cal A}$ from predictions to actions.

I assume that the user is an expected utility maximizer and holds a personal probability distribution $\pi(x,y)$ on the observables relevant for their decision problem. $\pi$ will affect the replicability analysis via the choice of the optimal decision function. This distribution may or may not reflect information arising in studies $S_1, \ldots, S_K$, but, to begin, will be independent of~$k$. While it is critical that studies $S_1, \ldots, S_K$ are not used by the modeler in the development of $\phi$, the same does not necessarily hold, in my view, for the user, although the nature of the replicability evaluation does change depending on whether $\pi$ reflects these studies.

An optimal decision function $\delta^*$ satisfies
$$
\delta^*(\phi) = \max _{\delta \in \Delta} E _{\pi} \left \{ U ( \delta(\phi(x)), x, y ) \right \}.
$$
I assume the user, as a rational agent, will utilize $\phi$ solely via $\delta^*$.

The assessor, for each study in turn, will describe the user's utility through the vectors 
$$ 
    \bigl( U ( \delta^*(\phi(x_{1k})), x_{1k}, y_{1k} ), \ldots, U ( \delta^*(\phi(x_{n_kk})), x_{n_kk}, y_{n_kk} ) \bigr)
$$
for $k=1, \ldots, K$. 
For study $k$, the user's utility is, on average, 
\begin{equation} \label{eq:uk}
{\cal U}_k = \frac 1 {n_k} \sum _{i=1} ^{n_k} U ( \delta^*(\phi(x_{ik})), x_{ik}, y_{ik} ).
\end{equation}

From here, I propose to define the prediction rule $\phi$ to be replicable if its optimal application to the same decision problem in different data sets leads to approximately the same average utility to the user. The degree of approximation can be formalized in many ways, and could itself be viewed as a decision problem if the assessor's role can be modeled in those terms. 

A summary of deviations among ${\cal U}_k$'s is a useful point of departure. For example, the $K \times K$ matrix 
$
{\cal U} 
$
with generic element ${\cal U}_k - {\cal U}_{k'}$ could be examined or visualized. 
Binary summarizations of the ${\cal U}_k$'s can be used to define replicability. Two examples are in the following definitions:
\begin{definition}[Absolute $\epsilon$-replicability] \label{def:er} \phantom{ } \\
$\phi$ is $\epsilon$-replicable in absolute utility over $S_1, \ldots, S_K$ if 
$$
\max _{k,k'} | {\cal U}_k - {\cal U}_{k'} | \leq \epsilon
$$
\end{definition}
\begin{definition}[Relative $\epsilon$-replicability]  \label{def:errel} \phantom{ } \\
$\phi$ is $\epsilon$-replicable in relative utility over $S_1, \ldots, S_K$ if 
$$
\max _{k,k'} \frac { 2 | {\cal U}_k - {\cal U}_{k'} | }{{\cal U}_k + {\cal U}_{k'}} \leq \epsilon
$$
\end{definition}

Definitions~\ref{def:er} and~\ref{def:errel} can be applied both to replication by design and observational replication, depending on how $S_1, \ldots, S_K$ is formed.

Definitions~\ref{def:er} and~\ref{def:errel} are conditional on observed data. Agreeing on the conclusions only requires agreeing on the choice of studies and data integrity.

This descriptive, empirical, definition is in contrast to potential definitions that may require additional theoretical constructs, such as collections of hypothetical datasets defined by a data generating model, or families of such data generating models. In such constructs, the summation in Equation~(\ref{eq:uk}) would be replaced by the expectation with respect to a joint predictive distribution reflecting the assessor's knowledge and beliefs. This distribution would not necessarily coincide with $\pi (x,y)$.

The ideas of this section are the basic building blocks for replicability assessment across multiple users and utility specifications. For example, our user could have multiple applications for the same prediction rule in different decision problems, each requiring a separate replicability analysis. More generally, the process could be repeated for various users separately, as illustrated below, each with potentially different decision spaces, priors, and utility functions. 


\subsection{Examples} \label{sec:ex-roles}

\subsubsection{VA Example.}\label{sec:va3}

In the VA study \citep{Lee2021dc}, the modelers are 5 participating companies which commercialize the algorithms considered. 
The replicability evaluation is carried out in parallel for 7 algorithms provided by the 5 companies.
To the extent that any of these algorithms focus on the same clinical detection task, the companies are in direct competition. The users are physicians in the VA Health System. It is unknown, but possible, that individual physicians may have dual interest with some of the companies. The assessors are scientists working within the VA system. The utility of the algorithms is defined for the VA as a collective. User and assessor in this case are somewhat aligned, but are not necessarily in complete agreement. Assessors report not to have duality of interest with the companies \citep{Lee2021dc}. This exemplifies the complex overlap of interests in the three roles. 

The study collection consists of $K=2$ VA hospitals, one in Seattle and the other in Atlanta. These two studies are used a) together, to produce a replicability-by-design analysis of previous claims, and b) separately, to examine replicability across components of the VA system. Labels are abstracted from medical records. A subset was regraded by a second expert, and differing grades were arbitrated by a retina specialist who did not know the identities of the graders. This illustrates the strengths gained from inter-subjective agreement on data. The clinical decision varies with the algorithm. For simplicity one may approximate it as whether or not additional follow-up is needed based on the retinal scan. Though multiple metrics are examined, the closest to a utility is the "value per encounter", defined as "the estimated pricing of each algorithm to make a normal profit (i.e., revenue and costs $= 0$) if deployed at the VA. This calculation was based on a two-stage scenario in which an AI algorithm would be used initially and then the images that screened negative would not need additional review by an optometrist or ophthalmologist" \cite[page 1170]{Lee2021dc}. In the replicability analysis $| {\cal U}_k - {\cal U}_{k'} |$ is the difference in value of encounter between Atlanta and Seattle. This turned out to be nontrivial, owing, according to the authors, both to differences in the populations served, and to the quality of the images. One of the centers did not perform a useful preliminary dilation as often as the other before collecting the images.

\subsubsection{Ovarian Cancer Example.} \label{ex:oc3}
In the Ovarian Cancer Study \citep{Waldron2014}, the modelers are 14 research groups who published prognostic algorithms meeting a set of criteria in terms of clinical goals and reproducibility of code. The replicability evaluation is carried out in parallel. The assessors are scientists funded by the NIH. There is some overlap between the assessors and the modelers for at least two of the models. The potential conflict is addressed by "freezing" the algorithms to the version originally published and by providing a transparent and reproducible analysis workflow for the replicability work. The potential users are physicians, although none of the 14 algorithm was in broad clinical use at the time of the evaluation. In fact the replicability analysis included among its goals to assess whether this family of rules was ready for clinical application. No formal decision framework is considered in \citep{Waldron2014}.

\subsection{Dominance}

\begin{figure}[b]
    \centering
    \includegraphics[width=.45\textwidth]{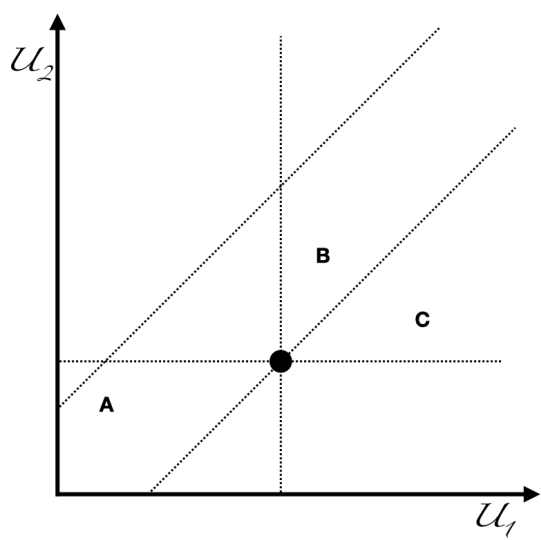}
    \caption{Regions of Average Utility for comparison between $\phi$ and alternative classifiers. The dot is positioned at coordinates ${\cal U}_1,{\cal U}_2$ for $\phi$. The two $45^{\circ}$ lines are equidistant from the main diagonal, not shown. Letters denote regions of interest considered in the text.}
    \label{fig:dom}
\end{figure}

Consider now comparing $\phi$ to other classifiers. Beginning with two studies, a useful perspective is to partition the average utility space as in Figure~\ref{fig:dom}. The dot is positioned at coordinates ${\cal U}_1,{\cal U}_2$ for $\phi$. Compared to $\phi$, an alternative classifier in region B would display a better average utility in both studies, as well as better replicability, because the empirical average utilities would be closer to each other. An alternative in region C displays better average utility in both studies, but worse replicability; the reverse is true in region A. 

This reasoning suggests that we can define dominance in this context as follows. Consider classifiers $\phi$ and $\phi'$, the latter with average utilities ${\cal U}'_{k}$, $k = 1, \ldots, K$.

\begin{definition}[Dominance] \label{def:dom}
$\phi$ dominates $\phi'$ in absolute utility and replicability over $S_1, \ldots, S_K$ if 
$$
\max _{k,k'} | {\cal U}_k - {\cal U}_{k'} | \leq \max _{k,k'} | {\cal U}'_k - {\cal U}'_{k'} |
$$
and
$$
{\cal U}'_k \leq {\cal U}_k \quad \quad k = 1, \ldots, K.
$$
\end{definition}

Formally, this definition could be applied to the context of Section~\ref{sec:dtr}. 
However, a decision maker with their own probability model $\pi$ may not decide based on the empirical average utilities, but rather update $\pi$ in the light of the studies and evaluate expected utilities accordingly. On the other hand, the assessor would remain interested in the empirical average utilities, and, to the extent that the utility function captures objectives of common interest, may value comparisons such as those of Figure~\ref{fig:dom}.

Figure~\ref{fig:dom} suggests the possibility of a formalization where the assessors hold their own utility function, capturing replicability. We could then leverage multi-agent decision theory approaches \cite{keen:raif:meye:1976} to understand the trade offs between the goals of accuracy and replicability. This might be interesting in some applications but too restrictive in others. For example, if $\phi$ is of general public health utility, both the user and the assessor would have an interest in high values of $\cal U$.

\section{The Binary Case}

\subsection{Replicability of Binary Prediction Rules}

In many applications, the product of the prediction rule are binary class labels rather that probabilities or risk scores. I will denote this special class of algorithms 
by the script variant of the letter phi, as in
$\varphi : \cal X \rightarrow \cal Y$. 

In the user-centered approach of Section~\ref{sec:dtr}
this is a special case obtained setting ${\cal A} = {\cal Y}$ and considering the rule $\varphi$ constructed from prediction/scoring rule $\phi$ via:
$$
\varphi (x) = \delta^*(\phi(x)) = \max _{\delta \in \Delta} 
E _{\pi} \left \{ U ( \delta(\phi(x)), x, y ) \right \}.
$$
The definition of ${\cal U}_k$ specializes to:
\begin{equation} \label{eq:ukc}
{\cal U}_k = \frac 1 {n_k} \sum _{i=1} ^{n_k} U ( \varphi(x_{ik})), x_{ik}, y_{ik} )
\end{equation}
based on which we can apply Definitions~\ref{def:er} and~\ref{def:errel}. 

A formal tie between $\varphi$ on one hand and the trio ($\delta$, $U$, $\pi$) on the other is not always made explicitly. Nonetheless it is often the case that binary prediction rules are built with some consideration of the modeler's expectation of $x$ and $y$, and desired properties of $\delta$ for users. A common example occurs when modelers first build a prediction rule $\phi$ and then apply thresholds to dichotomize the results. Another interesting scenario covered by this case is one where modeler and user are the same entity and the action space is binary.

When $\delta$ and $\pi$ are not explicitly specified, one can still evaluate replicability of $\varphi$ by positing a utility function directly as
$$U (\varphi, y) : ( \cal Y \times \cal X \times \cal Y) \rightarrow \mathbb{R}.$$
For example, if $U ( \varphi, x, y ) = I_{\varphi = y}$ for every $x$, then ${\cal U}_k$ is the empirical proportion of cases where prediction and outcomes coincide in study $k$ and $\epsilon$-replicability obtains when this proportion does not vary by more than $\epsilon$ in any two-study comparison.
Here replicability can be characterized without reference to a user's subjective model $\pi$. The utilities still refer to a specific, albeit hypothetical decision problem. In practice utilities like $U ( \varphi, x, y ) = I_{\varphi = y}$ are widely used in practice for simplicity and in the hope that they may capture the quality of a classifier well enough across several potential applications. 

Another commonly used approach in binary prediction is to separately penalize the two possible errors. This is particularly important in medical applications where it is rarely the case that the error of assigning a low risk individual to the high risk class is as severe as the opposite. This holds across a wide range of clinical applications.

A typical generalization of $U ( \varphi, x, y ) = I_{\varphi = y}$ is 
\begin{equation} \label{eq:u01c}
U ( \varphi, x, y ) = u_{01} I_{\varphi < y} + u_{10} I_{\varphi > y},
\end{equation}
where $u_{01}$ and $u_{10}$ are negative numbers quantifying the consequences of each of the two error types.
Defining the relative frequencies of the study-specific errors as
\[
f_{01}^k = \frac 1 {n_k} \sum _{i=1} ^{n_k} I _{ \phi(x_{ik}) < y_{ik} }
\]
and
\[
f_{10}^k = \frac 1 {n_k} \sum _{i=1} ^{n_k} I _{ \phi(x_{ik}) > y_{ik} }
\]
leads to
\begin{equation} \label{eq:uk01c}
{\cal U}_k = u_{01} f_{01}^k + u_{10} f_{10}^k.
\end{equation}
In the canonical two-by-two table of labels versus classifications, $f_{01}^k$ and $f_{10}$ represent the off-diagonal joint probabilities. 
Any, but not necessarily all, aspects of the $K$ joint distributions of $\phi$ and $y$ may be relevant for assessing replicability. The utility function reflects information on which are relevant in a specific problem. For example, the two frequencies of correct predictions are not distinguished in this utility specification, as they are both assigned the maximum utility, that is $0$. The function $U$ defines a three-set partition of the four-set sample space for $\phi$ and $y$ into equivalence classes relevant for the user's decision. In this framework, cross-study variation in the frequencies of correct predictions is irrelevant for user-based replicability as long as the overall proportion remains constant.

Expression~(\ref{eq:uk01c}) can be further rewritten in terms of the empirical sensitivity and specificity of detection of $y=1$, and the empirical proportions of $y=1$ cases, as long as all the latter are nonzero. Viewed this way, replicability requires sufficient stability of these three quantities across studies, so that their combination according to~(\ref{eq:uk01c}) may not vary by more than $\epsilon$. The empirical proportions of $y=1$ cases remains necessary for the evaluation of (\ref{eq:uk01c}). As a corollary, metrics that depend solely on sensitivity and specificity cannot be viewed as special cases of (\ref{eq:uk01c}).

Importantly, the same collection of studies may meet the user-centered replicability definition for one user and not another, depending on of their specification of $u_{01}$ and $u_{10}$. The same classifier may be sufficiently replicable for a user that requires high sensitivity, but not for another requiring high specificity. 

Definition~\ref{def:errel} can be used to build extensions to multiple utility functions. These could arise either from the need for the agent to use $\phi$ in multiple applications, or from the desire to define replicability across a set of agents, each with their own utility. 
Defining user-centered replicability over a class of users requires either scaling the utility functions so they are comparable or using a vector of bounds instead of a single $\epsilon$ the definition. This variant is not pursued here in any detail. 

\subsection{Examples}\label{sec:ex-binary}

\subsubsection{TB example.}
In \cite{Qin2019sr}, the modelers are companies developing algorithms for image analysis. The assessors are supported by philanthropic funding, and no conflicts with companies are reported. Assessors selected companies based on a comprehensive literature review, further strengthening the independence of the assessment. The user is a hypothetical TB clinician in one of the two countries considered, Cameroon and Nepal.

The algorithms generate ``abnormality scores'' $\phi$. Labels $y$ encode bacteriologically confirmed TB status. Assignment of images to predicted classes $\delta^* (\phi)$ is based on a threshold on the abnormality score. According to the authors, there are no generally recommended threshold scores to use, which motivates them to consider ROC curves. 

Each specification of $u_{01}$ and $u_{10}$ implies an optimal threshold, which depends on these quantities as well as empirical specificity, sensitivity and prevalence \cite{Metz1978snm}. Perhaps the ROC could be viewed as a step towards assessing performance and replicability across a range of specifications for $u_{01}$ and $u_{10}$, each corresponding to a different hypothetical user. However, explicit consideration of prevalence is required for the computation of any instance of~(\ref{eq:ukc}), and is lacking from ROC analysis and related summaries. A similar consideration applies to the C-index for time-to-event outcomes used in \citep{Waldron2014} in Example~\ref{ex:oc}.

\subsubsection{Sepsis.} 
In Example~\ref{ex:um}, $\phi$ produces a risk score used for classifying patients according to their risk of sepsis. The developer is a company, providing IT services to many hospitals. Again we have multiple users, each aiming to optimally use the score to inform clinical decisions in their emergency rooms. Assessors are academic researchers, some of whom work at one of the $K=24$ hospitals studied. 
Replicability is evaluated across hospitals as well as over time within each hospital. Replicability is evaluated with respect to the proportion of patients generating sepsis alerts per day, or $\sum_i I _{ \varphi(x_{ik}) = 1 }$.


\subsection{The case of K=1} \label{sec:k1}

In replicability by design, as well as in many model validation efforts \cite{Steyerberg2014ehj,Collins2014bmc} one may begin the assessment with a single study $S_1$. User-based replicability can still be carried out if benchmark value ${\cal U}_0$ is available, typically from the modeler. In this case replicability analysis reduces to the comparison of the vector 
$$ 
\bigl( \; U ( \varphi(x_{11}), x_{11}, y_{11} ), \ldots, 
        U ( \varphi(x_{n_11}), x_{n_11}, y_{n_11} ) \; \bigr)
$$
to the benchmark ${\cal U}_0$, followed by appropriate summarizations.
When knowledge of the user's $\delta^*$ is available to the modelers, they can then evaluate ${\cal U}_0$ using their own validation data, or earlier published evidence, or set-aside data from the training set. In another scenario, modeler and user cooperate in evaluating ${\cal U}_0$ before external assessment.

\section{Distance Replicability}

In this section I consider a scenario with a modeler and an assessor, possibly coinciding, and no user. This may be relevant in early stages of development of a model, before individual users are identified specifically, or in the case of models that target a very wide range of users with distinct goals, as do models embedded in smartphone apps.

Irrespective of the decision problem at hand and of the user's utility, the realized average utilities  ${\cal U}_1, \ldots, {\cal U}_K$ only depend on the data through the triplets 
$$(\phi(x_{ik})), x_{ik}, y_{ik}) \quad i = 1, \ldots n_k \quad k=1, \ldots, K.$$
Let $F_k$ be the empirical joint cumulative distribution of the points $(\phi(x_{ik})), x_{ik}, y_{ik})$ $i = 1, \ldots n_k$. A driver of replicability across different utility functions is the similarity among the distributions $F_1, \ldots, F_K$. If $\phi$ is a classifier for a discrete label, and the utility only depends on $x$ through $\phi$, then $F$ is a simple bivariate distribution on a contingency table, or confusion matrix.
If $\phi$ generates a probability distribution, more general spaces are required, but the concepts are the same. 

Mirroring the development in Section~\ref{sec:dtr}, define \\ $D(F_k,F_{k'})$ to be a distance between the c.d.f.'s $F_k$ and $F_{k'}$, such as the total variation distance on the appropriate space. Then we can posit the following:

\begin{definition}[Distance $\epsilon$-replicability] \label{def:erd}
$\phi$ is $\epsilon$-replicable in distance over $S_1, \ldots, S_K$ if 
$$
\max _{k,k'} D(F_k,F_{k'})  \leq \epsilon
$$
\end{definition}

A predictor $\phi$ defines a partition of ${\cal X}$ into sets with equal $\phi$. Generally, for given $S_1, \ldots, S_K$, the coarser the partition, the smaller $\max _{k,k'} D(F_k,F_{k'})$ will be. Even for the degenerate case in which $\phi$ is constant over ${\cal X}$, Definition~\ref{def:erd} may not hold owing to differences in the distribution of $y$'s.

A contrast with definitions~\ref{def:er} and~\ref{def:errel} is provided by the following observation.
For given $\pi$ and $U$, ${\cal U}_k$ is a functional of $F_k$. The difference 
$$ D_{\pi,U} (F_k,F_{k'}) \equiv | {\cal U}_k - {\cal U}_{k'}| $$
fails to satisfy the definition of distance among empirical c.d.f's, because it is possible to have $ D_{\pi,U} = 0$ with $F_k \neq F_{k'}$. 

Analyses that consider solely properties of the distributions of prediction rules conditional on class labels, such as the ROC curve or the C-index are also not covered by this definition, again because equality of conditional distributions alone does not imply equality of the joint distributions.

When $\epsilon = 0$, we can refer to distance replicability and user-based replicability as exact. This case is not of much practical interest as long as applicability is defined descriptively, as sampling variation will generally be present and will generate some variation across studies. 
Nonetheless it is conceptually interesting to note that exact distance replicability is a more strict requirement than exact user-based replicability. In other words if $\phi$ is exactly replicably by distance than it must be exactly replicable for any user in the sense of Section~\ref{sec:dtr}.

To see this, consider that equality of the empirical c.d.f's requires equality of the support points and associated point masses. This in turn occurs only if one of the two studies is formed by collating $b$ copies of the other, $b = 1, 2, 3, \ldots$. From this follows that each element in the sum~(\ref{eq:uk}) for the study with smaller sample size has $b$ identical terms in the sum~(\ref{eq:uk}) for the other. As the sample size of the larger studies is $b$ times that of the smaller one, $b$ cancels in the averaging and the result follows.

For an example, return to the setting of Expression~(\ref{eq:uk01c}) and define $f^k_{00}$ and $f^k_{11}$ to be the frequencies of the two possible correct classifications. A pair of studies such that $f^{k}_{00} + f^k_{11} = f^{k'}_{00} + f^{k'}_{11}$, $f^{k}_{01} = f^{k'}_{01}$, and $f^{k}_{10} = f^{k'}_{10}$ but $f^{k}_{00} \neq f^{k'}_{00}$ will have exact replicability for  users with utility~(\ref{eq:u01c}) but will fail to achieve exact distance replicability.

\section{Inference}

\subsection{Uncertainty quantification for replicability}

All the descriptive statements presented so far could be complemented by uncertainty quantification. Inevitably, this would require an additional layer of assumptions on the part of the assessor. I will mention here approaches that require a minimal amount of modeling and therefore a modest degree of additional stipulations. 

To begin, one can consider resampling. Bootstrap of units within each study in turn would provide variance estimates for each ${\cal U}_k$ and, assuming independence of the studies, of each element of the matrix ${\cal U}$. Variance estimates obtained in this way would condition on the selection of studies $S_1, \ldots, S_K$. For a simple extension,
Davison and Hinkley \cite{davison1997boostrap} describe a randomized cluster bootstrap procedure where both clusters (in this case studies) and observations within a cluster are sampled with replacement. 

In some cases, uncertainty may extend to study membership of individual units. In one example, data are extracted from a single encompassing data collection infrastructure and partitioned into $k$ studies based on geographical or administrative criteria, which could reasonably be specified at different levels of resolution. In another individuals may be assigned to studies based on ethnicity, a trait that may not be known with certainty in some cases. The study strap approach of Loewinger \etal \cite{Loewinger2022aas}, is a resampling technique that generates a collection of “pseudo-studies”, generalizing the randomized cluster bootstrap. The study strap is controlled by a tuning parameter that determines the proportion of observations to draw from each study, and can be used to dial the amount of study heterogeneity in the synthetic data throughout a range going from what is empirically observed to the case of complete exchangeability of units. The latter extreme is not a useful setting for a replicability analysis, but choosing tuning parameters that generate collections of studies close to the empirical distribution could provide a useful sensitivity analysis.

For a given $\epsilon$, and for any of the definitions in the preceding sections, resampling procedures would produce a proportion of cases that satisfy $\epsilon$-replicability. In turn, these could serve as an uncertainty quantification of whether $\phi$ meets the definition. 

\subsection{Rejecting Replicability} \label{sec:rej}

Another basic inferential question is whether replicability can be rejected via a significance testing approach. For an example with distance replicability, consider $\max _{k,k'} D(F_k,F_{k'})$ from Definition~(\ref{def:erd}) to be the test statistic of interest.  A simple procedure for producing significance statements in this context is to generate a permutation null for the vector $( {\cal U}_1 , \ldots, {\cal U}_K )$ and its functions by permutations of the study labels, and compute a p-value based on the permutation distribution. Multiple testing methods can also be relevant if one wishes to separately assess replicability for each of the pairwise comparisons. Elements of ${\cal U}$ are not independent, which requires additional care.

\section{Discussion}

\subsection{Sampling Frame} \label{sec:frame}

In my definition of a study, the sampling frame is the space $( \cal X \times \cal Y )$. Before units are sampled into a study, both $x$ and $y$ are unknown. All the descriptive measures of replicability proposed in the paper consider joint variation of both $x$ and $y$. Any heterogeneity of this joint variation across studies can and should challenge replicability. In machine learning terminology, replicability should be challenged by any of label shift, covariate shift or concept shift \citep{Kouw2019a, Zhuang2020a}.
This applies to both the utility-based definitions and the distance-based definitions. 

This theory does not cover efforts aiming at the useful but more limited goal of assessing the discrimination ability of prediction rules. These efforts, in analysis and often in design, condition on the class labels $y$. Examples include the ubiquitous ROC analysis which is often used as a criterion, or the sole criterion, for evaluating prediction rules. In scenarios with label shift, failures of replicability with important practical consequences can elude class-conditional analyses.

\subsection{Local Replicability}

My discussion considered properties of algorithms when applied to entire datasets. In this sense they are all global properties. In many applications it may be very interesting to consider groups within these studies. Replicability could be differentially evaluated within each group. If we define ${\cal X}^* \subset {\cal X}$ to be any subset of the feature space, we can revisit every definition given in the preceding sections, upon restricting the analysis to $x \in {\cal X}^*$, provided the set is not empty.
In general, the variation of ${\cal U}$ across studies will depend on the ${\cal X}^*$ chosen, and it may be the case that replicability is achieved in some groups but not others.
When ${\cal X}$ is coarse and cells are sufficiently populated, this logic can be pushed to the level of considering each cell separately to serve as ${\cal X}^*$. 


\subsection{Algorithmic Fairness}

It is interesting to think about replicability across collections of studies where individuals in different studies have the same rights (applicants for credit or for educational opportunities), or users of predictions have the same ethical responsibilities (medical providers). In this type of circumstance, it may be possible to construct meaningful collections of studies around notions of algorithmic fairness of the prediction rule studied \cite{Dwork2012acm,Zemel:2013wz}. Algorithmic fairness in classification is concerned with preventing discrimination against individuals based on their membership in some group. A connection arises with (global) distance replicability if one chooses the labels $1, \ldots, K$ to represent these groups. If $\phi$ satisfies distance replicability, it will be difficult for any user to discriminate among groups, on average, using $\phi$. 

It is more difficult to tie fairness to user-based replicability, as fair algorithms could produce different user's utilities in different groups, as a result of "benign" variation in the $F_k$'s, that is variation that is not associated with a discriminative use of $\phi$. 

Alternatively, a protected groups assignment could be used to investigate local replicability by appropriately choosing ${\cal X}^*$. Achieving local replicability within a protected group would not protect from discrimination if it exists, but may quantify its replicability.

These considerations apply to groups. Dwork \etal \cite{Dwork2012acm} define a seminal framework for fair classification at the individual level. In their words, it comprises:
{\it
\begin{enumerate}
    \item a (hypothetical) task-specific metric for determining the degree to which individuals are similar with respect to the classification task at hand;
    \item an algorithm for maximizing utility subject to the fairness constraint, that similar individuals are treated similarly.
\end{enumerate}
}
Here the definition of similarity among individuals should not include the group membership we intend to protect. Their approach, like the one I describe here, is also multi-agent, and also considers the modeler separately from the user. In contrast, it also explicitly considers the rights of the individuals being classified, which I did not consider. Extending the framework of this paper to include individuals as a fourth role could be interesting. Depending on the application context, individuals may have an interest in any subset of fairness, replicability, and prediction accuracy.

\subsection{Retraining}

I discussed how to define and quantify replicability of an algorithm $\phi$ that was previously trained and remains fixed throughout the analysis. My goal is to capture the implementation stage of a machine learning algorithm, once the development is completed. From a methodological perspective, however, the question of replicability can and should also be asked of model fitting techniques. 

Given a collection of $k$ studies, and analogously to what was described in Section~\ref{sec:dtr}, replicability of training techniques can be explored using designs that consider every pair of studies. For an example, in Bernau \etal\cite{Bernau2014} we define a cross-study validation matrix whose generic element measures predictive performance when one trains a predictor in study $k$ and evaluates it externally in $k'$. Our goal was to investigate properties of methodologies for training classifiers when external replicability is a goal. In contrast to how ${\cal U}$ is defined in Section~\ref{sec:dtr}, $\phi_k$ is different in every row and trained {\it de novo} using study $k$. Both ${\cal U}$ and the Bernau \etal version, offer the opportunity to learn about study heterogeneity and outlying studies. See also~\cite{Trippa2015b}.

Another useful design for investigating replicability of training techniques is the leave-one-study-out design \cite{riester2014}, which applies the jackknife logic at the study level. When training on $K-1$ studies, however, a challenge is to properly incorporate potential study-to-study heterogeneity, an issue considered in Section~\ref{sec:msl}.

In the social sciences, Vijayakumar and Cheung~\cite{Vijayakumar2021sscr} investigated the replication success of $R^2$ in both cross-validation and cross-study validation. They focus on three replication aims: 1) tests of inconsistency to examine whether single replications reject the originally reported study-specific $R^2$; 2) tests of consistency based on a region of equivalence, and 3) meta-analytic intervals for accuracy measures ---a goal also pursued by Waldron~\etal~\cite{Waldron2014}.
 
In addition to prediction performance and the utility thereof, it is interesting from a methodological standpoint to investigate replicability of various aspects of model construction, such as dimensionality, smoothness, variable selection and variable importance. 
Examples include~\cite{Vijayakumar2018zfp}. Yu and Kumbier~\cite{Yu2020pnas} emphasized model stability as a guiding principle. They define it as {\it acceptable consistency of a data result relative to appropriate perturbations of the data or model}. As examples of perturbation they suggest jackknife, bootstrap, and cross validation. Multi-study extensions would be interesting from a replicability standpoint.


\subsection{Testing Replicability and Replicability of Testing} \label{sec:rvalue}

In a testing approach to replicability of prediction rules, such as that sketched in Section~\ref{sec:rej}, the null hypothesis is the equality of the expected utility of a prediction rule across studies for a specific user or users. This has not to my knowledge been explored in depth.

On the other hand, there is a robust and useful literature on assessing the replicability of tests of hypotheses, which considers whether a hypothesis about the data generating mechanism is replicably rejected in multiple studies. Often this is framed in the context of multiple testing. An important foundational paper in this area is Heller and Benjamini~\citep{Heller:2014cn} who introduce the $r$-value, defined as the lowest false discovery rate at which a given finding can be called replicated. 

The context of meta-analysis, the systematic combination of results from studies investigating the same hypotheses, offers an interesting contrast to replicability analyses in terms of how the two approaches relate to study-to-study variation. For an anecdote tied to Example~\ref{ex:oc}, in \cite{Ganzfried2013} we identified expression of the gene CXCL12 as prognostic of overall survival in patients with ovarian cancer, via the combined analysis of 14 studies, in only two of which CXCL12 expression is a significant predictor. \citep{Jaljuli2022} offers additional and more systematic exploration of the discordance of goals and results between meta-analysis and replicability. 

Since the publication of~\cite{Ganzfried2013}, substantial biological evidence has accumulated on the prognostic role of this gene (see \cite{dalterio2022c} and references therein), thanks to progress in cancer immunology research. This supports the conclusion of the meta-analysis. On the other hand, the data of \citep{Ganzfried2013} would most likely provide evidence against the replicability of the hypothesis that CXCL12 expression is associated with survival, as measured by, say, an $r$-value. And although such analysis was not carried out, the data would likely have questioned replicability of prediction rules based solely on CXCL12 expression as well. Replicability reaches a different conclusion compared to meta-analysis because it asks a different question. Replicability is intentionally sensitive to the heterogeneity of study designs, the challenges in the normalization across technologies of the measured expression of a gene with only moderate transcriptional activity, and, if implemented via significance, the study sample sizes. In contrast, meta-analysis hopes to find signal in the midst of this variation. 

\subsection{Learning Replicability} \label{sec:msl}

After discovering failures of replicability, whether in inference or prediction, a reasonable next step is to move beyond a single study analysis, and tackle the study-to-study variation as part of the learning process. In inference, meta-analysis exemplifies this.

In prediction, availability of multiple datasets suitable to address the same or similar prediction question offer the opportunity to train algorithms that can incorporate knowledge of cross-study heterogeneity and produce predictions that are more likely be replicable in future data from the same or other studies. The domain generalization literature is particularly germane here as it focuses on leveraging multiple datasets in model training to improve prediction performance on an unseen, but related, domain \citep{Wang2021a}. 

In statistics, interest in drawing upon multiple data sets in prediction is also emerging. 
Approaches include meta-analyzing model coefficients (e.g. \cite{Riester:2012gt,Rashid2020jasa,Ventz2020a}) and ensembling models with weights that reward replicability. Specifically, in Patil and Parmigiani~\cite{Patil:2018hk}
we propose a multi-study generalization of stacking \citep{Breiman1996ml} to achieve this goal. Our approach comprises two stages: 1) training models on each study separately, and B) ensembling them via a stacking regression on the merged data. This structure rewards cross-study prediction performance as ensemble weights are primarily driven by how well each model predicts across studies different from the one where it was trained. 

\subsection{Prediction Tasks}

An interesting direction for generalization is to 
characterize replicability of a broadly defined prediction task, such as response to a drug treatment based on information on a patient's genome. In this case we would not necessarily have a specific $\phi$ or a methodology of interest. We would need to establish a class of measurements of $x$ and $y$ that constitute a sufficiently homogeneous collection to be worth studying from a replicability viewpoint, and potentially extend the definition of replicability to classes of $\phi$'s or optimally selected $\phi$'s within a class.

\subsection{Broader Perspectives}

An open question is whether or not to approach the definition of replicability from a game theoretic perspective. I imagine there would be many ways of meaningfully doing so depending on the context in which the prediction is developed, used, and assessed. My attempt here is to take a slightly more general perspective where the goal is to provide definitions that address trustworthiness of predictions across a relevant scientific community. Of course this is predicated on trust in the assessor and data. But even taking this trust for granted, there remains a gap before we arrive at statements about a specific prediction rule for a specific application. This is the gap my definitions try to fill.

While I discussed illustrative examples to make the concepts more concrete, I did not intend to provide a descriptive theory but rather to encourage discussion about prescriptive theories of replicability quantification. As a first step, my sense is that an explicit and transparent statement about who are the actors in the roles of modeler, user and assessor is a foundational step that should be encouraged in these analysis. The next is to explicitly connect the metrics used to quantify replicability to users' decisions.

I will close by noting that ultimately, as a field, we would benefit from examining the development, validation and implications of prediction rules in their historical, cultural, and social contexts. Such efforts would be close in scope to the field of science and technology studies \cite{Sismondo2004} which has already contributed important perspectives to epistemology. 

\begin{acks}[Acknowledgments]

I presented a preliminary version of Section~\ref{sec:acr} at a 2022 symposium on "Statistical methods and models for complex data", held in Padova. I am grateful to my discussants Marco Alf\`o and Gianmarco Alto\`e for very thoughtful comments, and to Aldo Solari for encouraging me to think about falsifiability in the context of replication. Lorenzo Trippa, Micheal Lavine and two insightful reviewers helped with comments on an earlier draft.

\end{acks}

\begin{funding}
Work supported by NSF-DMS 2113707.
\end{funding}



\bibliographystyle{imsart-number} 
\bibliography{mainArxiv1}       

\begin{thebibliography}{56}

\bibitem{Barba2018a}
\begin{barticle}[author]
\bauthor{\bsnm{{Barba}},~\bfnm{Lorena~A.}\binits{L.~A.}}
\btitle{{Terminologies for Reproducible Research}}.
\bpages{arXiv:1802.03311}.
\end{barticle}
\endbibitem

\bibitem{Beckers2021}
\begin{barticle}[author]
\bauthor{\bsnm{Beckers},~\bfnm{R.}\binits{R.}},
  \bauthor{\bsnm{Kwade},~\bfnm{Z.}\binits{Z.}} \AND
  \bauthor{\bsnm{Zanca},~\bfnm{F.}\binits{F.}}
\btitle{The EU medical device regulation: Implications for artificial
  intelligence-based medical device software in medical physics}.
\bvolume{83}
\bpages{1-8}.
\bdoi{https://doi.org/10.1016/j.ejmp.2021.02.011}
\end{barticle}
\endbibitem

\bibitem{Bernau2014}
\begin{barticle}[author]
\bauthor{\bsnm{Bernau},~\bfnm{Christoph}\binits{C.}},
  \bauthor{\bsnm{Riester},~\bfnm{Markus}\binits{M.}},
  \bauthor{\bsnm{Boulesteix},~\bfnm{Anne-Laure}\binits{A.-L.}},
  \bauthor{\bsnm{Parmigiani},~\bfnm{Giovanni}\binits{G.}},
  \bauthor{\bsnm{Huttenhower},~\bfnm{Curtis}\binits{C.}},
  \bauthor{\bsnm{Waldron},~\bfnm{Levi}\binits{L.}} \AND
  \bauthor{\bsnm{Trippa},~\bfnm{Lorenzo}\binits{L.}}
(\byear{2014}).
\btitle{Cross-study validation for the assessment of prediction algorithms.}
\bjournal{Bioinformatics}
\bvolume{30}
\bpages{i105--i112}.
\bnote{PMCID: PMC4058929}.
\bdoi{10.1093/bioinformatics/btu279}
\end{barticle}
\endbibitem

\bibitem{Breiman1996ml}
\begin{barticle}[author]
\bauthor{\bsnm{Breiman},~\bfnm{Leo}\binits{L.}}
(\byear{1996}).
\btitle{Stacked regressions}.
\bjournal{Machine Learning}
\bvolume{24}
\bpages{49--64}.
\bdoi{10.1007/BF00117832}
\end{barticle}
\endbibitem

\bibitem{Breiman2001}
\begin{barticle}[author]
\bauthor{\bsnm{Breiman},~\bfnm{Leo}\binits{L.}}
(\byear{2001}).
\btitle{{Statistical Modeling: The Two Cultures}}.
\bjournal{Statistical Science}
\bvolume{16}
\bpages{199--231}.
\bdoi{doi:10.1214/ss/1009213726}
\end{barticle}
\endbibitem

\bibitem{Association:98z9A4lW}
\begin{bbooklet}[author]
\bauthor{\bsnm{Broman},~\bfnm{Karl}\binits{K.}},
  \bauthor{\bsnm{Cetinkaya-Rundel},~\bfnm{Mine}\binits{M.}},
  \bauthor{\bsnm{Nussbaum},~\bfnm{Amy}\binits{A.}},
  \bauthor{\bsnm{Paciorek},~\bfnm{Christopher}\binits{C.}},
  \bauthor{\bsnm{Peng},~\bfnm{Roger}\binits{R.}},
  \bauthor{\bsnm{Turek},~\bfnm{Daniel}\binits{D.}}, \bauthor{} \AND
  \bauthor{\bsnm{Wickham},~\bfnm{Hadley}\binits{H.}}
\btitle{{Recommendations to Funding Agencies for Supporting Reproducible
  Research }}.
\end{bbooklet}
\endbibitem

\bibitem{Chang:2015ua}
\begin{barticle}[author]
\bauthor{\bsnm{Chang},~\bfnm{Lo-Bin}\binits{L.-B.}} \AND
  \bauthor{\bsnm{Geman},~\bfnm{Donald}\binits{D.}}
(\byear{2015}).
\btitle{{Tracking Cross-Validated Estimates of Prediction Error as Studies
  Accumulate}}.
\bjournal{Journal of the American Statistical Association}
\bvolume{110}
\bpages{1239--1247}.
\bdoi{10.1080/01621459.2014.1002926}
\end{barticle}
\endbibitem

\bibitem{Collins2014bmc}
\begin{barticle}[author]
\bauthor{\bsnm{Collins},~\bfnm{Gary~S.}\binits{G.~S.}}, \bauthor{\bparticle{de}
  \bsnm{Groot},~\bfnm{Joris~A.}\binits{J.~A.}},
  \bauthor{\bsnm{Dutton},~\bfnm{Susan}\binits{S.}},
  \bauthor{\bsnm{Omar},~\bfnm{Omar}\binits{O.}},
  \bauthor{\bsnm{Shanyinde},~\bfnm{Milensu}\binits{M.}},
  \bauthor{\bsnm{Tajar},~\bfnm{Abdelouahid}\binits{A.}},
  \bauthor{\bsnm{Voysey},~\bfnm{Merryn}\binits{M.}},
  \bauthor{\bsnm{Wharton},~\bfnm{Rose}\binits{R.}},
  \bauthor{\bsnm{Yu},~\bfnm{Ly-Mee}\binits{L.-M.}},
  \bauthor{\bsnm{Moons},~\bfnm{Karel~G.}\binits{K.~G.}} \AND
  \bauthor{\bsnm{Altman},~\bfnm{Douglas~G.}\binits{D.~G.}}
\btitle{External validation of multivariable prediction models: a systematic
  review of methodological conduct and reporting.}
\bvolume{14}
\bpages{40}.
\end{barticle}
\endbibitem

\bibitem{davison1997boostrap}
\begin{bbook}[author]
\bauthor{\bsnm{Davison},~\bfnm{Anthony~Christopher}\binits{A.~C.}} \AND
  \bauthor{\bsnm{Hinkley},~\bfnm{D.~V.}\binits{D.~V.}}
(\byear{1997}).
\btitle{Boostrap methods and their applications}.
\bpublisher{Cambridge University Press}, \baddress{New York}.
\end{bbook}
\endbibitem

\bibitem{Dwork2012acm}
\begin{binproceedings}[author]
\bauthor{\bsnm{Dwork},~\bfnm{Cynthia}\binits{C.}},
  \bauthor{\bsnm{Hardt},~\bfnm{Moritz}\binits{M.}},
  \bauthor{\bsnm{Pitassi},~\bfnm{Toniann}\binits{T.}},
  \bauthor{\bsnm{Reingold},~\bfnm{Omer}\binits{O.}} \AND
  \bauthor{\bsnm{Zemel},~\bfnm{Richard}\binits{R.}}
\btitle{Fairness through Awareness}.
In \bbooktitle{Proceedings of the 3rd Innovations in Theoretical Computer
  Science Conference}.
\bseries{ITCS '12}
\bpages{214–226}.
\bpublisher{Association for Computing Machinery}, \baddress{New York, NY, USA}.
\bdoi{10.1145/2090236.2090255}
\end{binproceedings}
\endbibitem

\bibitem{dalterio2022c}
\begin{barticle}[author]
\bauthor{\bsnm{D’Alterio},~\bfnm{Crescenzo}\binits{C.}},
  \bauthor{\bsnm{Spina},~\bfnm{Anna}\binits{A.}},
  \bauthor{\bsnm{Arenare},~\bfnm{Laura}\binits{L.}},
  \bauthor{\bsnm{Chiodini},~\bfnm{Paolo}\binits{P.}},
  \bauthor{\bsnm{Napolitano},~\bfnm{Maria}\binits{M.}},
  \bauthor{\bsnm{Galdiero},~\bfnm{Francesca}\binits{F.}},
  \bauthor{\bsnm{Portella},~\bfnm{Luigi}\binits{L.}},
  \bauthor{\bsnm{Simeon},~\bfnm{Vittorio}\binits{V.}},
  \bauthor{\bsnm{Signoriello},~\bfnm{Simona}\binits{S.}},
  \bauthor{\bsnm{Raspagliesi},~\bfnm{Francesco}\binits{F.}} \betal{et~al.}
(\byear{2022}).
\btitle{Biological Role of Tumor/Stromal CXCR4-CXCL12-CXCR7 in
  MITO16A/MaNGO-OV2 Advanced Ovarian Cancer Patients}.
\bjournal{Cancers}
\bvolume{14}
\bpages{1849}.
\end{barticle}
\endbibitem

\bibitem{Ebrahimian2022ar}
\begin{barticle}[author]
\bauthor{\bsnm{Ebrahimian},~\bfnm{Shadi}\binits{S.}},
  \bauthor{\bsnm{Kalra},~\bfnm{Mannudeep~K.}\binits{M.~K.}},
  \bauthor{\bsnm{Agarwal},~\bfnm{Sheela}\binits{S.}},
  \bauthor{\bsnm{Bizzo},~\bfnm{Bernardo~C.}\binits{B.~C.}},
  \bauthor{\bsnm{Elkholy},~\bfnm{Mona}\binits{M.}},
  \bauthor{\bsnm{Wald},~\bfnm{Christoph}\binits{C.}},
  \bauthor{\bsnm{Allen},~\bfnm{Bibb}\binits{B.}} \AND
  \bauthor{\bsnm{Dreyer},~\bfnm{Keith~J.}\binits{K.~J.}}
\btitle{FDA-regulated AI Algorithms: Trends, Strengths, and Gaps of Validation
  Studies}.
\bvolume{29}
\bpages{559-566}.
\bdoi{https://doi.org/10.1016/j.acra.2021.09.002}
\end{barticle}
\endbibitem

\bibitem{Finlayson2021nejm}
\begin{barticle}[author]
\bauthor{\bsnm{Finlayson},~\bfnm{Samuel~G.}\binits{S.~G.}},
  \bauthor{\bsnm{Subbaswamy},~\bfnm{Adarsh}\binits{A.}},
  \bauthor{\bsnm{Singh},~\bfnm{Karandeep}\binits{K.}},
  \bauthor{\bsnm{Bowers},~\bfnm{John}\binits{J.}},
  \bauthor{\bsnm{Kupke},~\bfnm{Annabel}\binits{A.}},
  \bauthor{\bsnm{Zittrain},~\bfnm{Jonathan}\binits{J.}},
  \bauthor{\bsnm{Kohane},~\bfnm{Isaac~S.}\binits{I.~S.}} \AND
  \bauthor{\bsnm{Saria},~\bfnm{Suchi}\binits{S.}}
\btitle{The Clinician and Dataset Shift in Artificial Intelligence}.
\bvolume{385}
\bpages{283-286}.
\bdoi{10.1056/NEJMc2104626}
\end{barticle}
\endbibitem

\bibitem{Fisher1925}
\begin{bbook}[author]
\bauthor{\bsnm{Fisher},~\bfnm{R.~A.}\binits{R.~A.}}
\btitle{Statistical methods for research workers}.
\bpublisher{Edinburgh Oliver \& Boyd}.
\end{bbook}
\endbibitem

\bibitem{Ganzfried2013}
\begin{barticle}[author]
\bauthor{\bsnm{Ganzfried},~\bfnm{Benjamin~Frederick}\binits{B.~F.}},
  \bauthor{\bsnm{Riester},~\bfnm{Markus}\binits{M.}},
  \bauthor{\bsnm{Haibe-Kains},~\bfnm{Benjamin}\binits{B.}},
  \bauthor{\bsnm{Risch},~\bfnm{Thomas}\binits{T.}},
  \bauthor{\bsnm{Tyekucheva},~\bfnm{Svitlana}\binits{S.}},
  \bauthor{\bsnm{Jazic},~\bfnm{Ina}\binits{I.}},
  \bauthor{\bsnm{Wang},~\bfnm{Xin~Victoria}\binits{X.~V.}},
  \bauthor{\bsnm{Ahmadifar},~\bfnm{Mahnaz}\binits{M.}},
  \bauthor{\bsnm{Birrer},~\bfnm{Michael~J}\binits{M.~J.}},
  \bauthor{\bsnm{Parmigiani},~\bfnm{Giovanni}\binits{G.}},
  \bauthor{\bsnm{Huttenhower},~\bfnm{Curtis}\binits{C.}} \AND
  \bauthor{\bsnm{Waldron},~\bfnm{Levi}\binits{L.}}
(\byear{2013}).
\btitle{curated{O}varian{D}ata: clinically annotated data for the ovarian
  cancer transcriptome.}
\bjournal{Database (Oxford)}
\bvolume{2013}
\bpages{bat013}.
\bnote{PMCID: PMC3625954}.
\bdoi{10.1093/database/bat013}
\end{barticle}
\endbibitem

\bibitem{Geisser2014wol}
\begin{barticle}[author]
\bauthor{\bsnm{Geisser},~\bfnm{Seymour}\binits{S.}}
\btitle{Predictive Analysis}.
\bjournal{Wiley StatsRef: Statistics Reference Online}.
\end{barticle}
\endbibitem

\bibitem{Goodman2016stm}
\begin{barticle}[author]
\bauthor{\bsnm{Goodman},~\bfnm{Steven~N.}\binits{S.~N.}},
  \bauthor{\bsnm{Fanelli},~\bfnm{Daniele}\binits{D.}} \AND
  \bauthor{\bsnm{Ioannidis},~\bfnm{John P.~A.}\binits{J.~P.~A.}}
\btitle{What does research reproducibility mean?}
\bvolume{8}
\bpages{341ps12--341ps12}.
\end{barticle}
\endbibitem

\bibitem{Heller:2014cn}
\begin{barticle}[author]
\bauthor{\bsnm{Heller},~\bfnm{Ruth}\binits{R.}},
  \bauthor{\bsnm{Bogomolov},~\bfnm{Marina}\binits{M.}} \AND
  \bauthor{\bsnm{Benjamini},~\bfnm{Yoav}\binits{Y.}}
(\byear{2014}).
\btitle{{Deciding whether follow-up studies have replicated findings in a
  preliminary large-scale omics study.}}
\bjournal{Proceedings of the National Academy of Sciences of the United States
  of America}
\bvolume{111}
\bpages{16262--16267}.
\bdoi{10.1073/pnas.1314814111}
\end{barticle}
\endbibitem

\bibitem{Jaljuli2022}
\begin{barticle}[author]
\bauthor{\bsnm{Jaljuli},~\bfnm{Iman}\binits{I.}},
  \bauthor{\bsnm{Benjamini},~\bfnm{Yoav}\binits{Y.}},
  \bauthor{\bsnm{Shenhav},~\bfnm{Liat}\binits{L.}},
  \bauthor{\bsnm{Panagiotou},~\bfnm{Orestis~A.}\binits{O.~A.}} \AND
  \bauthor{\bsnm{Heller},~\bfnm{Ruth}\binits{R.}}
\btitle{Quantifying Replicability and Consistency in Systematic Reviews}.
\bvolume{0}
\bpages{1-14}.
\bdoi{10.1080/19466315.2022.2050291}
\end{barticle}
\endbibitem

\bibitem{keen:raif:meye:1976}
\begin{bbook}[author]
\bauthor{\bsnm{Keeney},~\bfnm{Ralph~L}\binits{R.~L.}},
  \bauthor{\bsnm{Raiffa},~\bfnm{Howard~Auth}\binits{H.~A.}} \AND
  \bauthor{\bsnm{Meyer},~\bfnm{Richard F~Contributor}\binits{R.~F.~C.}}
(\byear{1976}).
\btitle{{Decisions With Multiple Objectives: Preferences and Value Tradeoffs}}.
\end{bbook}
\endbibitem

\bibitem{Kelly2019mbcm}
\begin{barticle}[author]
\bauthor{\bsnm{Kelly},~\bfnm{Christopher~J.}\binits{C.~J.}},
  \bauthor{\bsnm{Karthikesalingam},~\bfnm{Alan}\binits{A.}},
  \bauthor{\bsnm{Suleyman},~\bfnm{Mustafa}\binits{M.}},
  \bauthor{\bsnm{Corrado},~\bfnm{Greg}\binits{G.}} \AND
  \bauthor{\bsnm{King},~\bfnm{Dominic}\binits{D.}}
\btitle{Key challenges for delivering clinical impact with artificial
  intelligence}.
\bvolume{17}
\bpages{195}.
\bdoi{10.1186/s12916-019-1426-2}
\end{barticle}
\endbibitem

\bibitem{Kenett:2015kb}
\begin{barticle}[author]
\bauthor{\bsnm{Kenett},~\bfnm{Ron~S}\binits{R.~S.}} \AND
  \bauthor{\bsnm{Shmueli},~\bfnm{Galit}\binits{G.}}
(\byear{2015}).
\btitle{{Clarifying the terminology that describes scientific
  reproducibility}}.
\bjournal{Nature methods}
\bvolume{12}
\bpages{699--699}.
\bdoi{10.1038/nmeth.3489}
\end{barticle}
\endbibitem

\bibitem{Koh2020pmlr}
\begin{barticle}[author]
\bauthor{\bsnm{{Koh}},~\bfnm{Pang~Wei}\binits{P.~W.}},
  \bauthor{\bsnm{{Sagawa}},~\bfnm{Shiori}\binits{S.}},
  \bauthor{\bsnm{{Marklund}},~\bfnm{Henrik}\binits{H.}},
  \bauthor{\bsnm{{Xie}},~\bfnm{Sang~Michael}\binits{S.~M.}},
  \bauthor{\bsnm{{Zhang}},~\bfnm{Marvin}\binits{M.}},
  \bauthor{\bsnm{{Balsubramani}},~\bfnm{Akshay}\binits{A.}},
  \bauthor{\bsnm{{Hu}},~\bfnm{Weihua}\binits{W.}},
  \bauthor{\bsnm{{Yasunaga}},~\bfnm{Michihiro}\binits{M.}},
  \bauthor{\bsnm{{Lanas Phillips}},~\bfnm{Richard}\binits{R.}},
  \bauthor{\bsnm{{Gao}},~\bfnm{Irena}\binits{I.}},
  \bauthor{\bsnm{{Lee}},~\bfnm{Tony}\binits{T.}},
  \bauthor{\bsnm{{David}},~\bfnm{Etienne}\binits{E.}},
  \bauthor{\bsnm{{Stavness}},~\bfnm{Ian}\binits{I.}},
  \bauthor{\bsnm{{Guo}},~\bfnm{Wei}\binits{W.}},
  \bauthor{\bsnm{{Earnshaw}},~\bfnm{Berton~A.}\binits{B.~A.}},
  \bauthor{\bsnm{{Haque}},~\bfnm{Imran~S.}\binits{I.~S.}},
  \bauthor{\bsnm{{Beery}},~\bfnm{Sara}\binits{S.}},
  \bauthor{\bsnm{{Leskovec}},~\bfnm{Jure}\binits{J.}},
  \bauthor{\bsnm{{Kundaje}},~\bfnm{Anshul}\binits{A.}},
  \bauthor{\bsnm{{Pierson}},~\bfnm{Emma}\binits{E.}},
  \bauthor{\bsnm{{Levine}},~\bfnm{Sergey}\binits{S.}},
  \bauthor{\bsnm{{Finn}},~\bfnm{Chelsea}\binits{C.}} \AND
  \bauthor{\bsnm{{Liang}},~\bfnm{Percy}\binits{P.}}
\btitle{{WILDS: A Benchmark of in-the-Wild Distribution Shifts}}.
\bpages{arXiv:2012.07421}.
\end{barticle}
\endbibitem

\bibitem{Kouw2019a}
\begin{barticle}[author]
\bauthor{\bsnm{Kouw},~\bfnm{W.}\binits{W.}} \AND
  \bauthor{\bsnm{Loog},~\bfnm{M.}\binits{M.}}
(\byear{2019}).
\btitle{An introduction to domain adaptation and transfer learning}.
\bjournal{arXiv:1812.11806}.
\end{barticle}
\endbibitem

\bibitem{Lee2021dc}
\begin{barticle}[author]
\bauthor{\bsnm{Lee},~\bfnm{Aaron~Y.}\binits{A.~Y.}},
  \bauthor{\bsnm{Yanagihara},~\bfnm{Ryan~T.}\binits{R.~T.}},
  \bauthor{\bsnm{Lee},~\bfnm{Cecilia~S.}\binits{C.~S.}},
  \bauthor{\bsnm{Blazes},~\bfnm{Marian}\binits{M.}},
  \bauthor{\bsnm{Jung},~\bfnm{Hoon~C.}\binits{H.~C.}},
  \bauthor{\bsnm{Chee},~\bfnm{Yewlin~E.}\binits{Y.~E.}},
  \bauthor{\bsnm{Gencarella},~\bfnm{Michael~D.}\binits{M.~D.}},
  \bauthor{\bsnm{Gee},~\bfnm{Harry}\binits{H.}},
  \bauthor{\bsnm{Maa},~\bfnm{April~Y.}\binits{A.~Y.}},
  \bauthor{\bsnm{Cockerham},~\bfnm{Glenn~C.}\binits{G.~C.}},
  \bauthor{\bsnm{Lynch},~\bfnm{Mary}\binits{M.}} \AND
  \bauthor{\bsnm{Boyko},~\bfnm{Edward~J.}\binits{E.~J.}}
\btitle{{Multicenter, Head-to-Head, Real-World Validation Study of Seven
  Automated Artificial Intelligence Diabetic Retinopathy Screening Systems}}.
\bvolume{44}
\bpages{1168-1175}.
\bdoi{10.2337/dc20-1877}
\end{barticle}
\endbibitem

\bibitem{Leek2010nrg}
\begin{barticle}[author]
\bauthor{\bsnm{Leek},~\bfnm{Jeffrey~T.}\binits{J.~T.}},
  \bauthor{\bsnm{Scharpf},~\bfnm{Robert~B.}\binits{R.~B.}},
  \bauthor{\bsnm{Bravo},~\bfnm{Héctor~Corrada}\binits{H.~C.}},
  \bauthor{\bsnm{Simcha},~\bfnm{David}\binits{D.}},
  \bauthor{\bsnm{Langmead},~\bfnm{Benjamin}\binits{B.}},
  \bauthor{\bsnm{Johnson},~\bfnm{W.~Evan}\binits{W.~E.}},
  \bauthor{\bsnm{Geman},~\bfnm{Donald}\binits{D.}},
  \bauthor{\bsnm{Baggerly},~\bfnm{Keith}\binits{K.}} \AND
  \bauthor{\bsnm{Irizarry},~\bfnm{Rafael~A.}\binits{R.~A.}}
\btitle{Tackling the widespread and critical impact of batch effects in
  high-throughput data}.
\bvolume{11}
\bpages{733--739}.
\bdoi{10.1038/nrg2825}
\end{barticle}
\endbibitem

\bibitem{Leek2008}
\begin{barticle}[author]
\bauthor{\bsnm{Leek},~\bfnm{Jeffrey~T}\binits{J.~T.}} \AND
  \bauthor{\bsnm{Storey},~\bfnm{John~D}\binits{J.~D.}}
(\byear{2007}).
\btitle{{Capturing Heterogeneity in Gene Expression Studies by Surrogate
  Variable Analysis}}.
\bjournal{PLoS Genetics}
\bvolume{3}
\bpages{e161}.
\bdoi{10.1371/journal.pgen.0030161}
\end{barticle}
\endbibitem

\bibitem{Lemay2022njpdm}
\begin{barticle}[author]
\bauthor{\bsnm{Lemay},~\bfnm{Andreanne}\binits{A.}},
  \bauthor{\bsnm{Hoebel},~\bfnm{Katharina}\binits{K.}},
  \bauthor{\bsnm{Bridge},~\bfnm{Christopher~P.}\binits{C.~P.}},
  \bauthor{\bsnm{Befano},~\bfnm{Brian}\binits{B.}},
  \bauthor{\bsnm{De~Sanjosé},~\bfnm{Silvia}\binits{S.}},
  \bauthor{\bsnm{Egemen},~\bfnm{Didem}\binits{D.}},
  \bauthor{\bsnm{Rodriguez},~\bfnm{Ana~Cecilia}\binits{A.~C.}},
  \bauthor{\bsnm{Schiffman},~\bfnm{Mark}\binits{M.}},
  \bauthor{\bsnm{Campbell},~\bfnm{John~Peter}\binits{J.~P.}} \AND
  \bauthor{\bsnm{Kalpathy-Cramer},~\bfnm{Jayashree}\binits{J.}}
\btitle{Improving the repeatability of deep learning models with Monte Carlo
  dropout.}
\bvolume{5}
\bpages{174}.
\end{barticle}
\endbibitem

\bibitem{Loewinger2022aas}
\begin{barticle}[author]
\bauthor{\bsnm{Loewinger},~\bfnm{Gabriel}\binits{G.}},
  \bauthor{\bsnm{Patil},~\bfnm{Prasad}\binits{P.}},
  \bauthor{\bsnm{Kishida},~\bfnm{Kenneth~T.}\binits{K.~T.}} \AND
  \bauthor{\bsnm{Parmigiani},~\bfnm{Giovanni}\binits{G.}}
(\byear{2022}).
\btitle{{Hierarchical resampling for bagging in multistudy prediction with
  applications to human neurochemical sensing}}.
\bjournal{The Annals of Applied Statistics}
\bvolume{16}
\bpages{2145 -- 2165}.
\bdoi{10.1214/21-AOAS1574}
\end{barticle}
\endbibitem

\bibitem{Metz1978snm}
\begin{barticle}[author]
\bauthor{\bsnm{Metz},~\bfnm{Charles~E.}\binits{C.~E.}}
\btitle{Basic principles of ROC analysis}.
\bvolume{8}
\bpages{283-298}.
\bdoi{https://doi.org/10.1016/S0001-2998(78)80014-2}
\end{barticle}
\endbibitem

\bibitem{MorenoTorres2012pr}
\begin{barticle}[author]
\bauthor{\bsnm{Moreno-Torres},~\bfnm{Jose~G.}\binits{J.~G.}},
  \bauthor{\bsnm{Raeder},~\bfnm{Troy}\binits{T.}},
  \bauthor{\bsnm{Alaiz-Rodríguez},~\bfnm{Rocío}\binits{R.}},
  \bauthor{\bsnm{Chawla},~\bfnm{Nitesh~V.}\binits{N.~V.}} \AND
  \bauthor{\bsnm{Herrera},~\bfnm{Francisco}\binits{F.}}
\btitle{A unifying view on dataset shift in classification}.
\bvolume{45}
\bpages{521-530}.
\bdoi{https://doi.org/10.1016/j.patcog.2011.06.019}
\end{barticle}
\endbibitem

\bibitem{InstituteofMedicine:vr}
\begin{bbook}[author]
\bauthor{\bsnm{{Institute of Medicine}}}
\btitle{{Evolution of Translational Omics}}.
\bpublisher{iom.edu}.
\bdoi{10.17226/13297}
\end{bbook}
\endbibitem

\bibitem{CommitteeonReproducibilityandReplicabilityinScience:2019hx}
\begin{bbook}[author]
\bauthor{\bsnm{{Committee on Reproducibility and Replicability in Science}}}
(\byear{2019}).
\btitle{{Reproducibility and Replicability in Science}}.
\bpublisher{National Academies Press}, \baddress{Washington, D.C.}
\bdoi{10.17226/25303}
\end{bbook}
\endbibitem

\bibitem{Patil:2018hk}
\begin{barticle}[author]
\bauthor{\bsnm{Patil},~\bfnm{Prasad}\binits{P.}} \AND
  \bauthor{\bsnm{Parmigiani},~\bfnm{Giovanni}\binits{G.}}
(\byear{2018}).
\btitle{{Training replicable predictors in multiple studies}}.
\bjournal{Proceedings of the National Academy of Science, USA}
\bvolume{115}
\bpages{2578--2583}.
\end{barticle}
\endbibitem

\bibitem{Qin2019sr}
\begin{barticle}[author]
\bauthor{\bsnm{Qin},~\bfnm{Zhi~Zhen}\binits{Z.~Z.}},
  \bauthor{\bsnm{Sander},~\bfnm{Melissa~S.}\binits{M.~S.}},
  \bauthor{\bsnm{Rai},~\bfnm{Bishwa}\binits{B.}},
  \bauthor{\bsnm{Titahong},~\bfnm{Collins~N.}\binits{C.~N.}},
  \bauthor{\bsnm{Sudrungrot},~\bfnm{Santat}\binits{S.}},
  \bauthor{\bsnm{Laah},~\bfnm{Sylvain~N.}\binits{S.~N.}},
  \bauthor{\bsnm{Adhikari},~\bfnm{Lal~Mani}\binits{L.~M.}},
  \bauthor{\bsnm{Carter},~\bfnm{E.~Jane}\binits{E.~J.}},
  \bauthor{\bsnm{Puri},~\bfnm{Lekha}\binits{L.}},
  \bauthor{\bsnm{Codlin},~\bfnm{Andrew~J.}\binits{A.~J.}} \AND
  \bauthor{\bsnm{Creswell},~\bfnm{Jacob}\binits{J.}}
\btitle{Using artificial intelligence to read chest radiographs for
  tuberculosis detection: A multi-site evaluation of the diagnostic accuracy of
  three deep learning systems}.
\bvolume{9}
\bpages{15000}.
\bdoi{10.1038/s41598-019-51503-3}
\end{barticle}
\endbibitem

\bibitem{rams:1926}
\begin{bbook}[author]
\bauthor{\bsnm{Ramsey},~\bfnm{F}\binits{F.}}
(\byear{1926}).
\btitle{{The Foundations of Mathematics}}.
\end{bbook}
\endbibitem

\bibitem{Rashid2020jasa}
\begin{barticle}[author]
\bauthor{\bsnm{Rashid},~\bfnm{Naim~U}\binits{N.~U.}},
  \bauthor{\bsnm{Li},~\bfnm{Quefeng}\binits{Q.}},
  \bauthor{\bsnm{Yeh},~\bfnm{Jen~Jen}\binits{J.~J.}} \AND
  \bauthor{\bsnm{Ibrahim},~\bfnm{Joseph~G}\binits{J.~G.}}
(\byear{2020}).
\btitle{{Modeling Between-Study Heterogeneity for Improved Replicability in
  Gene Signature Selection and Clinical Prediction}}.
\bjournal{Journal of the American Statistical Association}
\bvolume{115}
\bpages{1125-1138}.
\bnote{PMID: 33012902}.
\bdoi{10.1080/01621459.2019.1671197}
\end{barticle}
\endbibitem

\bibitem{Riester:2012gt}
\begin{barticle}[author]
\bauthor{\bsnm{Riester},~\bfnm{Markus}\binits{M.}},
  \bauthor{\bsnm{Taylor},~\bfnm{Jennifer~M}\binits{J.~M.}},
  \bauthor{\bsnm{Feifer},~\bfnm{Andrew}\binits{A.}},
  \bauthor{\bsnm{Koppie},~\bfnm{Theresa}\binits{T.}},
  \bauthor{\bsnm{Rosenberg},~\bfnm{Jonathan~E}\binits{J.~E.}},
  \bauthor{\bsnm{Downey},~\bfnm{Robert~J}\binits{R.~J.}},
  \bauthor{\bsnm{Bochner},~\bfnm{Bernard~H}\binits{B.~H.}} \AND
  \bauthor{\bsnm{Michor},~\bfnm{Franziska}\binits{F.}}
(\byear{2012}).
\btitle{{Combination of a novel gene expression signature with a clinical
  nomogram improves the prediction of survival in high-risk bladder cancer.}}
\bjournal{Clinical Cancer Research}
\bvolume{18}
\bpages{1323--1333}.
\bdoi{10.1158/1078-0432.CCR-11-2271}
\end{barticle}
\endbibitem

\bibitem{Riester:2014ga}
\begin{barticle}[author]
\bauthor{\bsnm{Riester},~\bfnm{Markus}\binits{M.}},
  \bauthor{\bsnm{Wei},~\bfnm{Wei}\binits{W.}},
  \bauthor{\bsnm{Waldron},~\bfnm{Levi}\binits{L.}},
  \bauthor{\bsnm{Culhane},~\bfnm{Aedin~C}\binits{A.~C.}},
  \bauthor{\bsnm{Trippa},~\bfnm{Lorenzo}\binits{L.}},
  \bauthor{\bsnm{Oliva},~\bfnm{Esther}\binits{E.}},
  \bauthor{\bsnm{Kim},~\bfnm{Sung-Hoon}\binits{S.-H.}},
  \bauthor{\bsnm{Michor},~\bfnm{Franziska}\binits{F.}},
  \bauthor{\bsnm{Huttenhower},~\bfnm{Curtis}\binits{C.}},
  \bauthor{\bsnm{Parmigiani},~\bfnm{Giovanni}\binits{G.}} \AND
  \bauthor{\bsnm{Birrer},~\bfnm{Michael~J}\binits{M.~J.}}
(\byear{2014}).
\btitle{{Risk prediction for late-stage ovarian cancer by meta-analysis of 1525
  patient samples.}}
\bjournal{JNCI Journal of the National Cancer Institute}
\bvolume{106}
\bpages{dju048--dju048}.
\bdoi{10.1093/jnci/dju048}
\end{barticle}
\endbibitem

\bibitem{riester2014}
\begin{barticle}[author]
\bauthor{\bsnm{Riester},~\bfnm{Markus}\binits{M.}},
  \bauthor{\bsnm{Wei},~\bfnm{Wei}\binits{W.}},
  \bauthor{\bsnm{Waldron},~\bfnm{Levi}\binits{L.}},
  \bauthor{\bsnm{Culhane},~\bfnm{Aedin~C}\binits{A.~C.}},
  \bauthor{\bsnm{Trippa},~\bfnm{Lorenzo}\binits{L.}},
  \bauthor{\bsnm{Oliva},~\bfnm{Esther}\binits{E.}},
  \bauthor{\bsnm{Kim},~\bfnm{Sung-Hoon}\binits{S.-H.}},
  \bauthor{\bsnm{Michor},~\bfnm{Franziska}\binits{F.}},
  \bauthor{\bsnm{Huttenhower},~\bfnm{Curtis}\binits{C.}},
  \bauthor{\bsnm{Parmigiani},~\bfnm{Giovanni}\binits{G.}} \AND
  \bauthor{\bsnm{Birrer},~\bfnm{Michael~J}\binits{M.~J.}}
(\byear{2014}).
\btitle{Risk Prediction for Late-Stage Ovarian Cancer by Meta-analysis of 1525
  Patient Samples.}
\bjournal{J Natl Cancer Inst}.
\bdoi{10.1093/jnci/dju048}
\end{barticle}
\endbibitem

\bibitem{sava:1954}
\begin{bbook}[author]
\bauthor{\bsnm{Savage},~\bfnm{Leonard~J}\binits{L.~J.}}
(\byear{1954}).
\btitle{{The foundations of statistics}}.
\bpublisher{Wiley}, \baddress{New York}.
\end{bbook}
\endbibitem

\bibitem{Sismondo2004}
\begin{bbook}[author]
\bauthor{\bsnm{Sismondo},~\bfnm{Sergio}\binits{S.}}
(\byear{2004}).
\btitle{An introduction to science and technology studies}.
\bpublisher{Blackwell}.
\end{bbook}
\endbibitem

\bibitem{Steyerberg2014ehj}
\begin{barticle}[author]
\bauthor{\bsnm{Steyerberg},~\bfnm{Ewout~W.}\binits{E.~W.}} \AND
  \bauthor{\bsnm{Vergouwe},~\bfnm{Yvonne}\binits{Y.}}
\btitle{{Towards better clinical prediction models: seven steps for development
  and an ABCD for validation}}.
\bvolume{35}
\bpages{1925-1931}.
\bdoi{10.1093/eurheartj/ehu207}
\end{barticle}
\endbibitem

\bibitem{Stigler1982jrssa}
\begin{barticle}[author]
\bauthor{\bsnm{Stigler},~\bfnm{Stephen~M.}\binits{S.~M.}}
\btitle{Thomas Bayes's Bayesian Inference}.
\bvolume{145}
\bpages{250--258}.
\end{barticle}
\endbibitem

\bibitem{Trippa2015b}
\begin{barticle}[author]
\bauthor{\bsnm{Trippa},~\bfnm{Lorenzo}\binits{L.}},
  \bauthor{\bsnm{Waldron},~\bfnm{Levi}\binits{L.}},
  \bauthor{\bsnm{Huttenhower},~\bfnm{Curtis}\binits{C.}} \AND
  \bauthor{\bsnm{Parmigiani.},~\bfnm{Giovanni}\binits{G.}}
(\byear{2015}).
\btitle{Bayesian nonparametric cross-study validation of prediction methods.}
\bjournal{Ann. Appl. Stat.}
\bvolume{9}
\bpages{402-428}.
\end{barticle}
\endbibitem

\bibitem{Ventz2020a}
\begin{barticle}[author]
\bauthor{\bsnm{Ventz},~\bfnm{Steffen}\binits{S.}},
  \bauthor{\bsnm{Mazumder},~\bfnm{Rahul}\binits{R.}} \AND
  \bauthor{\bsnm{Trippa},~\bfnm{Lorenzo}\binits{L.}}
(\byear{2020}).
\btitle{Integration of Survival Data from Multiple Studies}.
\end{barticle}
\endbibitem

\bibitem{Vijayakumar2021sscr}
\begin{barticle}[author]
\bauthor{\bsnm{Vijayakumar},~\bfnm{Ranjith}\binits{R.}} \AND
  \bauthor{\bsnm{Cheung},~\bfnm{Mike W.~L.}\binits{M.~W.~L.}}
\btitle{Assessing Replicability of Machine Learning Results: An Introduction to
  Methods on Predictive Accuracy in Social Sciences}.
\bvolume{39}
\bpages{768--801}.
\end{barticle}
\endbibitem

\bibitem{Vijayakumar2018zfp}
\begin{barticle}[author]
\bauthor{\bsnm{Vijayakumar},~\bfnm{Ranjith}\binits{R.}} \AND
  \bauthor{\bsnm{Cheung},~\bfnm{Mike W.~L.}\binits{M.~W.~L.}}
(\byear{2018}).
\btitle{Replicability of machine learning models in the social sciences: A case
  study in variable selection.}
\bjournal{Zeitschrift f{\"u}r Psychologie}
\bvolume{226}
\bpages{259 - 273}.
\end{barticle}
\endbibitem

\bibitem{Waldron2014}
\begin{barticle}[author]
\bauthor{\bsnm{Waldron},~\bfnm{Levi}\binits{L.}},
  \bauthor{\bsnm{Haibe-Kains},~\bfnm{Benjamin}\binits{B.}},
  \bauthor{\bsnm{Culhane},~\bfnm{Aed{\`i}n~C}\binits{A.~C.}},
  \bauthor{\bsnm{Riester},~\bfnm{Markus}\binits{M.}},
  \bauthor{\bsnm{Ding},~\bfnm{Jie}\binits{J.}},
  \bauthor{\bsnm{Wang},~\bfnm{Xin~Victoria}\binits{X.~V.}},
  \bauthor{\bsnm{Ahmadifar},~\bfnm{Mahnaz}\binits{M.}},
  \bauthor{\bsnm{Tyekucheva},~\bfnm{Svitlana}\binits{S.}},
  \bauthor{\bsnm{Bernau},~\bfnm{Christoph}\binits{C.}},
  \bauthor{\bsnm{Risch},~\bfnm{Thomas}\binits{T.}},
  \bauthor{\bsnm{Ganzfried},~\bfnm{Benjamin~Frederick}\binits{B.~F.}},
  \bauthor{\bsnm{Huttenhower},~\bfnm{Curtis}\binits{C.}},
  \bauthor{\bsnm{Birrer},~\bfnm{Michael}\binits{M.}} \AND
  \bauthor{\bsnm{Parmigiani},~\bfnm{Giovanni}\binits{G.}}
(\byear{2014}).
\btitle{Comparative Meta-analysis of Prognostic Gene Signatures for Late-Stage
  Ovarian Cancer.}
\bjournal{J Natl Cancer Inst}
\bvolume{106}
\bpages{dju049}.
\bdoi{10.1093/jnci/dju049}
\end{barticle}
\endbibitem

\bibitem{Wang2021a}
\begin{barticle}[author]
\bauthor{\bsnm{Wang},~\bfnm{J.}\binits{J.}},
  \bauthor{\bsnm{Lan},~\bfnm{C.}\binits{C.}},
  \bauthor{\bsnm{Liu},~\bfnm{C.}\binits{C.}},
  \bauthor{\bsnm{Ouyang},~\bfnm{Y.}\binits{Y.}} \AND
  \bauthor{\bsnm{Qin},~\bfnm{T.}\binits{T.}}
(\byear{2021}).
\btitle{Generalizing to Unseen Domains: {A} Survey on Domain Generalization}.
\bjournal{arXiv:2103.03097}.
\end{barticle}
\endbibitem

\bibitem{Wong2021hi}
\begin{barticle}[author]
\bauthor{\bsnm{Wong},~\bfnm{Andrew}\binits{A.}},
  \bauthor{\bsnm{Cao},~\bfnm{Jie}\binits{J.}},
  \bauthor{\bsnm{Lyons},~\bfnm{Patrick~G.}\binits{P.~G.}},
  \bauthor{\bsnm{Dutta},~\bfnm{Sayon}\binits{S.}},
  \bauthor{\bsnm{Major},~\bfnm{Vincent~J.}\binits{V.~J.}},
  \bauthor{\bsnm{Ötleş},~\bfnm{Erkin}\binits{E.}} \AND
  \bauthor{\bsnm{Singh},~\bfnm{Karandeep}\binits{K.}}
\btitle{{Quantification of Sepsis Model Alerts in 24 US Hospitals Before and
  During the COVID-19 Pandemic}}.
\bvolume{4}
\bpages{e2135286-e2135286}.
\bdoi{10.1001/jamanetworkopen.2021.35286}
\end{barticle}
\endbibitem

\bibitem{Wu2021nm}
\begin{barticle}[author]
\bauthor{\bsnm{Wu},~\bfnm{Eric}\binits{E.}},
  \bauthor{\bsnm{Wu},~\bfnm{Kevin}\binits{K.}},
  \bauthor{\bsnm{Daneshjou},~\bfnm{Roxana}\binits{R.}},
  \bauthor{\bsnm{Ouyang},~\bfnm{David}\binits{D.}},
  \bauthor{\bsnm{Ho},~\bfnm{Daniel~E}\binits{D.~E.}} \AND
  \bauthor{\bsnm{Zou},~\bfnm{James}\binits{J.}}
\btitle{How medical AI devices are evaluated: limitations and recommendations
  from an analysis of FDA approvals}.
\bvolume{27}
\bpages{582--584}.
\end{barticle}
\endbibitem

\bibitem{Yu2020pnas}
\begin{barticle}[author]
\bauthor{\bsnm{{Yu}},~\bfnm{Bin}\binits{B.}} \AND
  \bauthor{\bsnm{{Kumbier}},~\bfnm{Karl}\binits{K.}}
\btitle{{Veridical data science}}.
\bvolume{117}
\bpages{3920-3929}.
\bdoi{10.1073/pnas.1901326117}
\end{barticle}
\endbibitem

\bibitem{Zemel:2013wz}
\begin{binproceedings}[author]
\bauthor{\bsnm{Zemel},~\bfnm{Richard}\binits{R.}},
  \bauthor{\bsnm{Swersky},~\bfnm{Kevin}\binits{K.}},
  \bauthor{\bsnm{Pitassi},~\bfnm{Toniann}\binits{T.}} \AND
  \bauthor{\bsnm{Dwork},~\bfnm{Cynthia}\binits{C.}}
(\byear{2013}).
\btitle{{Learning fair representations}}.
In \bbooktitle{Proceedings of the 30th International Conference on Machine
  Learning}.
\end{binproceedings}
\endbibitem

\bibitem{Zhang2020b}
\begin{barticle}[author]
\bauthor{\bsnm{Zhang},~\bfnm{Yuqing}\binits{Y.}},
  \bauthor{\bsnm{Patil},~\bfnm{Prasad}\binits{P.}},
  \bauthor{\bsnm{Johnson},~\bfnm{W~Evan}\binits{W.~E.}} \AND
  \bauthor{\bsnm{Parmigiani},~\bfnm{Giovanni}\binits{G.}}
(\byear{2020}).
\btitle{{Robustifying Genomic Classifiers To Batch Effects Via Ensemble
  Learning}}.
\bjournal{Bioinformatics}.
\bnote{btaa986}.
\bdoi{10.1093/bioinformatics/btaa986}
\end{barticle}
\endbibitem

\bibitem{Zhuang2020a}
\begin{barticle}[author]
\bauthor{\bsnm{Zhuang},~\bfnm{F.}\binits{F.}},
  \bauthor{\bsnm{Qi},~\bfnm{Z.}\binits{Z.}},
  \bauthor{\bsnm{Duan},~\bfnm{K.}\binits{K.}},
  \bauthor{\bsnm{Xi},~\bfnm{D.}\binits{D.}},
  \bauthor{\bsnm{Zhu},~\bfnm{Y.}\binits{Y.}},
  \bauthor{\bsnm{Zhu},~\bfnm{H.}\binits{H.}},
  \bauthor{\bsnm{Xiong},~\bfnm{H.}\binits{H.}} \AND
  \bauthor{\bsnm{He},~\bfnm{Q.}\binits{Q.}}
(\byear{2020}).
\btitle{A Comprehensive Survey on Transfer Learning}.
\bjournal{arXiv:02685}.
\end{barticle}
\endbibitem

\end{thebibliography}

\end{document}